\documentclass[12pt]{revtex4}

\begin{document}


\title{Schr\"{o}dinger Equation in Rotating Frame by using Stochastic Variational Method}

\author{Tomoi Koide}

\affiliation{IF, Universidade Federal do Rio de Janeiro, C.P.
68528, 21941-972, Rio de Janeiro, Brazil}

\author{Kazuo Tsushima}

\affiliation{
LFTC Universidade Cruzeiro do Sul, Rua Galvao Bueno, 868 , Liberdade
01506-000, S\~{a}o Paulo, 
Brazil}

\author{Takeshi Kodama}
\affiliation{IF, Universidade Federal do Rio de Janeiro, C.P.
68528, 21941-972, Rio de Janeiro, Brazil}
%
%

\begin{abstract}
We give a pedagogical introduction of the stochastic variational method by considering   
the quantization of a non-inertial particle system. We show that the effects of fictitious forces 
are represented in the forms of vector fields which behave analogous to the gauge fields 
in the electromagnetic interaction. We further discuss that the operator expressions for observables  
can be defined by applying the stochastic Noether theorem.
\end{abstract}

\maketitle

\section{Introduction}
\label{sec:int}

Variational approach conceptually plays a fundamental role in elucidating the structure of
classical mechanics, clarifying the origin of dynamics and the relation between symmetries and
conservation laws. 
On the other hand, its operations to classical and quantum systems lack coherence. 
In fact, the Lagrangian in classical mechanics is usually given by $T-V$ where 
$T$ and $V$ are a kinetic and a potential terms, respectively, but that 
for the derivation of the Schr\"{o}dinger 
equation, it does not have such a structure. 
That is, any clear or direct correspondence between classical and quantum mechanics 
does not seem to exist in the variational point of view.

However, if we extend the idea of the variation to stochastic
variable, the variational principle can describe classical and quantum behaviors in a unified way.
This method is called stochastic variational method (SVM) and firstly proposed by Yasue
\cite{yasue,guerra,pavon,nagasawa,cresson} 
so as to reformulate Nelson's stochastic quantization~\cite{nelson}. This framework is,
however, based on special techniques attributed to stochastic calculus which is not familiar to
physicists.

In this paper, we introduce this method by applying to the quantization of 
a non-inertial particle system, which is still controversial.
The appearance of the nontrivial interference effect of wave functions on the
rotating non-inertial frame was experimentally observed in 1979~\cite{werner}. 
Later Sakurai pointed out that
such an effect can be understood through the similarity between the Coriolis force and the
Lorentz force~\cite{sakurai}. 
So far, there are various approaches to derive the Schr\"{o}dinger equation 
in the non-inertial frame~\cite{mash, anandan, takagi, klink,kamebuchi}.

\vspace{-0.5cm}

\section{Classical equation in non-inercial system}
\label{sec:classical}

Let us introduce a non-inertial frame on which the position is denoted by ${\bf q}$.
Expressing the position in an inertial frame by ${\bf r}$, the transformation of these vectors 
are defined by 
\begin{eqnarray}
{\bf q} = {\bf R}(t) {\bf r} + {\bf c}(t), \label{co-tra}
\end{eqnarray}
where ${\bf c}(t)$ is a time-dependent translation, and ${\bf R}(t)$ is a general $3 \times 3$ 
rotation matrix satisfying 
${\bf R}^T (t){\bf R}(t) = 1$ .
Both of ${\bf r}$ and ${\bf q}$ are given by the Cartesian coordinate.

We consider usually one particle system in the inertial frame. Applying the coordinate transform by 
Eq.\ (\ref{co-tra}), the same system observed in the non-inertial frame is characterized by the 
following Lagrangian, 
\begin{eqnarray}
L 
&=& 
\frac{M}{2} ( \dot{\bf q} + {\bf A}({\bf q},t) + {\bf B}(t))^{2} - V({\bf q}), \label{cal-lag} 
\end{eqnarray}
where $V$ is the potential and, we introduce vector fields as 
\begin{eqnarray}
{\bf A}({\bf q},t) = {\bf R} \dot{\bf R}^{T} ({\bf q}-{\bf c}), \ \ \ \ 
{\bf B} (t) = -\dot{\bf c}.
\end{eqnarray}
The equations of motion obtained from this Lagrangian are given by 
\begin{eqnarray}
{\bf p} 
&=& M \left( \dot{\bf q} + {\bf A}({\bf q},t) + {\bf B}(t) \right), \\
\partial_t {\bf p}^i
&=& ({\bf R} \dot{\bf R}^T)^{ji} {\bf p}^j  - \partial_i V ({\bf q}). \label{eq-p}
\end{eqnarray}

\vspace{-0.5cm}

\section{Stochastic Variational Method}
\label{subsec:svm}

The discussion in this section follows the pedagogical introduction of SVM given by Ref.~\cite{dice2014}.
For the alternative review, see Ref.~\cite{zam}.

In the variational principle for stochastic variables, 
a particle trajectory is not any more smooth, and given by a zig-zag path in general.
As the consequence, the evolution of a particle trajectory is defined by the following 
forward stochastic differential equation (SDE),
\begin{equation}
d{\bf q}(t) = \left( \frac{{\bf p}({\bf q}(t),t)}{M} - {\bf A}({\bf q},t) - {\bf B}(t) \right)dt 
+ \sqrt{2\nu}d{\bf W}_t~~~~(dt > 0). \label{fsde}
\end{equation}
Here ${\bf p}({\bf x},t)$ is an unknown field determined by the stochastic variation.
Note that in the following ${\bf x}$ is used to denote the spatial parameter in the non-inertial frame.
The last term in Eq.\ (\ref{fsde}) is the origin of the zig-zag motion and called noise term. 
The parameter $\nu$ characterizes the strength of this noise term. 
The property of ${\bf W}_t$ is given by the standard Wiener process, which is characterized
by the following correlation properties,
\begin{eqnarray}
E[d{\bf W}_t] &=& 0, \ \ \ 
E[(dW^i_t)(dW^j_t)] = |dt| \delta^{ij},~~(i,j = x,y,z), \label{corr2}\\
E[W^i_t dW^j_{t'}] &=& 0~~{\rm for}~~(t\le t'),
\end{eqnarray}
where $E[~~~]$ indicates the average of stochastic events.

The probabilistic nature of the particle distribution described by Eq.\ (\ref{fsde}) 
is easily characterized by introducing the probability distribution defined by 
$\rho({\bf q},t) = \int d^3 {\bf q}_i~\rho_I ({\bf q}_i) E[ \delta^{(3)}({\bf q} - {\bf q}(t)) ]$,
where ${\bf q}(t)$ (more exactly ${\bf q}(t;{\bf q}_i)$) is the solution of Eq.~(\ref{fsde}) 
and $\rho_I ({\bf q}_i)$ is the initial particle distribution at an initial time $t_i$.
As is well-known, the evolution equation of $\rho({\bf q},t)$ is calculated form the SDE (\ref{fsde}) 
and is called the Fokker-Planck equation,
\begin{equation}
\partial_t \rho ({\bf x},t) 
= \nabla \cdot \left\{ - \left( \frac{{\bf p}({\bf x},t)}{M} - {\bf A}({\bf x},t) - {\bf B}(t) \right)
+ \nu \nabla \right\} \rho ({\bf x},t). \label{ffp}
\end{equation}

If the probability distribution evolves from 
$\rho_I ({\bf q})$ to $\rho_F ({\bf q}) \equiv \rho({\bf q}(t_f),t_f)$ at a final time $t_f$ following
Eq.\ (\ref{ffp}), the corresponding time-reversed process should describe 
the evolution from $\rho_F$ to $\rho_I$.
Suppose that this process is described by the backward SDE, 
\begin{equation}
d{\bf q}(t) = \left( \frac{\tilde{\bf p}({\bf q}(t),t)}{M} - {\bf A}({\bf q},t) - {\bf B}(t) \right)dt 
+ \sqrt{2\nu}d{\bf W}_t,~~~~(dt < 0). \label{bsde}
\end{equation}
To reproduce Eq.\ (\ref{ffp}) from the backward SDE, 
we find that the following consistency condition should be satisfied, 
${\bf p}({\bf x},t) = \tilde{\bf p}({\bf x},t) + 2\nu \nabla \ln \rho({\bf x},t)$.

We should stress that the usual definition of the particle velocity is not applicable, because 
$d\hat{\bf r}/dt$ is not well defined in the vanishing limit of $dt$ due to the singular behavior of ${\bf W}_t$.
The possible time differential in such a case was studied by Nelson \cite{nelson} and it is known that there are two possibilities: One is the the mean forward derivative
\begin{equation}
D {\bf q}(t) = \lim_{dt \rightarrow 0+} E \left[ \frac{{\bf q}(t + dt) - {\bf q}(t)}{dt} \Big| {\cal P}_t \right],
\end{equation}
and the other the mean backward derivative,
\begin{equation}
\tilde{D} {\bf q}(t) = \lim_{dt \rightarrow 0-} E \left[ \frac{{\bf q}(t + dt) - {\bf q}(t)}{dt} \Big| {\cal F}_t \right].
\end{equation}
These expectations are conditional averages, where ${\cal P}_t$ (${\cal F}_t$) indicates to fix values of ${\bf r}(t')$ for $t' \le t~~(t' \ge t)$.
For the $\sigma$-algebra of all measurable events of ${\bf r}(t)$,
$\{\mathcal{P}_{t}\}$ and $\{\mathcal{F}_{t}\}$ represent an increasing and a
decreasing family of sub-$\sigma$-algebras, respectively.
Applying these to Eqs.\ (\ref{fsde}) and (\ref{bsde}), we obtain, respectively,  
\begin{eqnarray}
D {\bf q}(t) = \frac{{\bf p}({\bf q},t)}{M} - {\bf A}({\bf q},t) -{\bf B}(t), 
~~~~ \tilde{D} {\bf q}(t) &=& \frac{\tilde{\bf p}({\bf q},t)}{M} - {\bf A}({\bf q},t) 
-{\bf B}(t).
\end{eqnarray}

\vspace{-0.5cm}

\section{Quantization in non-inertial frame}

Let us apply the stochastic variation to the system given by the Lagrangian (\ref{cal-lag}).
Then the particle trajectory in Eq.\ (\ref{cal-lag}) should be replaced by the 
stochastic one as was discussed in the previous section. 
Due to the existence of the two definitions of the time derivatives $D$ and $\tilde{D}$,
there is an ambiguity for the replacement of the kinetic term. 
In this work, we adopt the following replacement, 
\begin{eqnarray}
L ({\bf q}, D{\bf q}, \tilde{D}{\bf q}) 
= \frac{m}{2} 
\left[ \frac{(D{\bf q}(t) + {\bf A} + {\bf B})^2 
+ (\tilde{D}{\bf q}(t) + {\bf A} + {\bf B})^2 }{2}
\right]
- V({\bf q}(t)).
\label{sto-lag}
\end{eqnarray}
See Ref.\ \cite{koide1} for more precise discussion of this replacement.

The stochastic variation of the particle Lagrangian leads to the stochastic Euler-Lagrange equation, 
\begin{eqnarray}
\left. \tilde{D}\frac{\partial L}{\partial (D{\bf q}(t))} 
+ {D}\frac{\partial L}{\partial (\tilde{D}{\bf q}(t))} 
- \frac{\partial L}{\partial {\bf q}(t)}
\right|_{{\bf q}(t) = {\bf x}} = 0.
\end{eqnarray}
Here ${\bf q}(t)$ is replaced by the position parameter ${\bf x}$ at the last step of the calculation.
Substituting Eq.\ (\ref{sto-lag}), we obtain 
\begin{equation}
\left(\partial_t + \left(\frac{{\bf p}_m}{M}  - {\bf A} -{\bf B}\right)\cdot \nabla \right) {\bf p}_m 
-2M\nu^2 \nabla \rho^{-1/2} \Delta \sqrt{\rho}
= 
{\bf p}_m  \cdot \nabla_i {\bf A} 
- \nabla_i V, 
\end{equation}
where ${\bf p}_m = ({\bf p} + \tilde{\bf p})/2$.

This result of the variation can be re-expressed in the form of the Schr\"{o}dinger equation 
by introducing the wave function defined by 
$\Psi ({\bf x},t) = \sqrt{\rho({\bf x},t)}e^{i\theta({\bf x},t)}$.
Here $\rho({\bf x},t)$ is the probability distribution introduced above Eq. (\ref{ffp}), 
and the phase $\theta({\bf x},t)$ is defined by 
${\bf p}_m = 2M \nu \nabla \theta ({\bf x},t) $. 
Then we find that the evolution equation of the wave function is given by the following Schr\"{o}dinger 
equation,  
\begin{eqnarray}
i\hbar \partial_t \Psi ({\bf x},t)
= \left[
\frac{1}{2M} (-i\hbar \nabla - M ({\bf A}({\bf x},t) + {\bf B}(t)) )^2 
- \frac{M}{2} ({\bf A}({\bf x},t) + {\bf B}(t))^2 + V ({\bf x})
\right] \Psi({\bf x},t). \nonumber \\ \label{non-ine-sch}
\end{eqnarray}
Here we choose $\nu = \hbar/(2M)$.
One can see that the effect of the non-inertial forces appears through the vector fields ${\bf A}({\bf x},t)$ 
and ${\bf B}(t)$ which behaves as if the gauge potential in the electromagnetic interaction.

\vspace{-0.5cm}

\section{Observables}

The dynamics described by the above Schr\"{o}dinger equation satisfies Eherenfest's theorem.
In fact, the time evolution of the expectation value of the operator $-\hbar \nabla$ is given by 
\begin{eqnarray}
 \partial_t \langle -i\hbar \partial_i \rangle 
=
\langle ({\bf R} \dot{\bf R}^T)^{ji} (-i\hbar \partial_j) \rangle - \langle \partial_i V \rangle.
\end{eqnarray}
One can see that, if we can interpret $\hat{p} = -i\hbar \nabla$, the above equation 
corresponds to Eq.\ (\ref{eq-p}).

However, exactly speaking, it is not trivial whether we can interpret $-\hbar \nabla$ as the 
momentum operator even in the non-inertial frame. 
In SVM, the operator representations of observables are defined 
through the conservation laws obtained from the stochastic Lagrangian (\ref{sto-lag}).

For the sake of simplicity, let us consider the rotation around the z-axis, where
\begin{eqnarray}
 {\bf R}(t) =
 \left( 
 \begin{array}{ccc}
\cos \phi(t) & \sin \phi(t) & 0 \\
-\sin \phi (t) & \cos \phi (t) & 0 \\
0 & 0 & 1
 \end{array}
 \right), \ \ \ 
 {\bf c}(t) = 0.
\end{eqnarray}
This non-inertial system still holds the invariance for the rotation if $V({\bf x}) = V(|{\bf x}|)$. 
Then from the invariance of the stochastic action, we can obtain the angular momentum conservation of 
the present non-inertial system. 
For the infinitesimal rotation, ${\bf q}(t)$ is transformed as
$ {\bf q}(t) \longrightarrow {\bf q}(t) + {\bf A}(\phi(t))$, 
where ${\bf A}(\phi(t)) = \delta \dot{\phi} (-y, x, 0)$.

On the other hand, if the action is invariant for the above rotation, 
we can show the following quantity is conserved by 
applying the stochastic Noether theorem \cite{misawa,koide2}, 
\begin{eqnarray}
Q = E\left[ 
{\bf q}(t) \times 
\left( 
\frac{\partial L}{\partial (D{\bf q}(t))} 
+ 
\frac{\partial L}{\partial (\tilde{D}{\bf q}(t))}
\right)
\right].
\end{eqnarray}
Here $\times$ denotes the vector product.
Substituting the result of the stochastic variation, 
the above equation is re-expressed as
\begin{eqnarray}
Q = \int d^3 {\bf x}\ \ \Psi ({\bf x},t) 
{L}_z \Psi({\bf x},t),
\end{eqnarray}
where the angular momentum operator is introduced, 
$ {L}_z = -i\hbar (x \partial_y - y \partial_x)$.
This result means that $-i\hbar \nabla$ can be interpreted as the momentum operator even in the non-inertial 
system.

\vspace{-0.5cm}

\section{Concluding remarks}

We gave a brief summary of the stochastic variational method and 
showed that this is applicable to the quantization of the non-inertial particle system.
Then we find that the Eherenfest's theorem is still satisfied even for the Schr\"{o}dinger equation 
in the non-inertial frame and thus the result is consistent with those in Refs.~\cite{mash,anandan,takagi}, but 
different from Refs.~\cite{klink,kamebuchi}.

The advantage of the present approach compared to Refs.\ \cite{mash,anandan,takagi} is that 
the operator representations for observables are systematically obtained by applying the stochastic 
Noether theorem.

Although the framework of SVM was originally proposed to reformulate Nelson's stochastic quantization, 
its applicability is not restricted to quantization.
The derivation of the classical dissipative dynamics can be cast into the form of SVM: 
the Navier-Stokes-Fourier equation is obtain by employing the stochastic variation 
to the classical action of the Euler (ideal fluid) equation.
See Refs.~\cite{koide2,koide3} for details.

\vspace{1cm}

This work is financilly supported by CNPq.


\begin{thebibliography}{10}

\bibitem{yasue}
K. Yasue, J. Funct. Anal. \textbf{41}, 327 (1981).

\bibitem{guerra}
F. Guerra and L. M. Morato, Phys. Rev. D\textbf{27}, 1774 (1983).

\bibitem{pavon}
M. Pavon, J. Math. Phys. \textbf{36}, 6774 (1995).

\bibitem{nagasawa} 
M. Nagasawa, \textit{Stochastic Process in Quantum Physics} (Birkh\"{a}user, Basel, 2000).

\bibitem{cresson} J. Cresson and S. Darses, J. Math. Phys. \textbf{48}, 072703 (2007).

\bibitem{nelson}
E. Nelson, Phys. Rev. \textbf{150}, 1079 (1966).

\bibitem{werner}
S. A. Werner, J.-L. Staudenmann and R. Colella, Phys. Rev. Lett. \textbf{42}, 1103 (1979).

\bibitem{sakurai}
J. J. Sakurai, Phys. Rev. D\textbf{21}, 2993 (1980).

\bibitem{mash}
B. Mashhoon, Phys. Rev. Lett. \textbf{61}, 2639 (1988); \textbf{68}, 3812 (1992).

\bibitem{anandan}
J, Anandan and J. Suzuki, 
in \textit{Relativity in Rotating Frames: Relativistic Physics in Rotating Reference Frames}, 
ed. by G. Rizzi and M.L. Ruggiero. 
Fundamental Theories of Physics, vol 135 (Kluwer, Dordewcht, 2004) P361, 
arXiv:quant-ph/0305081v2.

\bibitem{takagi}
S. Takagi. Prog. Theor. Phys. \textbf{85}, 463 (1991).

\bibitem{klink}
W. H. Klink and S. Wickramasekara, Phys. Rev. Lett. \textbf{111}, 160404 (2013). 

\bibitem{kamebuchi}
S. Kamebuchi and M. Omote, Special Lecture on Quantum Mechanics, (Asakura, Tokyo, 2003) in Japanese.


\bibitem{dice2014}
T. Koide, T. Kodama and K. Tsishima, 
J. Phys.: Conf. Ser. \textbf{626} 012055 (2015).

\bibitem{zam}
J. C. Zambrini, Int. J. Theor. Phys.\textbf{24} 277 (1985).

\bibitem{koide1}
T. Koide, J. Phys.: Conf. Ser. \textbf{410} 012025 (2013).

\bibitem{misawa}
T. Misawa, J. Math. Phys. \textbf{29} 2178 (1988).

\bibitem{koide2}
T. Koide and T. Kodama, Prog. Theor. Exp. Phys. \textbf{093A03} (2015).

\bibitem {koide3}T. Koide and T. Kodama, J. Phys. A: Math. Theor. \textbf{45}, 255204 (2012).

\bibitem{koide4}
T. Koide, Phys. Lett. A\textbf{379}, 2007 (2015).


\end{thebibliography}
\end{document}